%% file: NewSingleBoson.tex
\documentstyle[12pt,epsfig,axodraw]{article}
\advance\textheight by \topskip
\def\pr#1 #2 #3 { {\rm Phys. Rev.}            {\bf #1}   (#2) #3}
\def\prd#1 #2 #3{ {\rm Phys. Rev.}            {\bf D#1}  (#2) #3}
\def\prl#1 #2 #3{ {\rm Phys. Rev. Lett.}      {\bf #1}   (#2) #3}
\def\plb#1 #2 #3{ {\rm Phys. Lett.}           {\bf B#1}  (#2) #3}
\def\npb#1 #2 #3{ {\rm Nucl. Phys.}           {\bf B#1}  (#2) #3}
\def\prp#1 #2 #3{ {\rm Phys. Rep.}            {\bf #1}   (#2) #3}
\def\zpc#1 #2 #3{ {\rm Z. Phys.}              {\bf C#1}  (#2) #3}
\def\epjc#1 #2 #3{ {\rm Eur. Phys. J.}        {\bf C#1}  (#2) #3}
\def\mpl#1 #2 #3{ {\rm Mod. Phys. Lett.}      {\bf A#1}  (#2) #3}
\def\ijmp#1 #2 #3{{\rm Int. J. Mod. Phys.}    {\bf A#1}  (#2) #3}
\def\ptp#1 #2 #3{ {\rm Prog. Theor. Phys.}    {\bf #1}   (#2) #3}
\def\jhep#1 #2 #3{ {\rm J. High Energy Phys.} {\bf #1}   (#2) #3}
\def\jphg#1 #2 #3{ {\rm J. Phys.}             {\bf G#1}  (#2) #3}
\def\cpc#1 #2 #3{ {\rm Comput. Phys. Commun.} {\bf #1}   (#2) #3}

\newcommand{\be}{\begin{equation}}
\newcommand{\ee}{\end{equation}}
\newcommand{\br}{\begin{eqnarray}}
\newcommand{\er}{\end{eqnarray}}
\newcommand{\ba}{\begin{array}}
\newcommand{\ea}{\end{array}}
\newcommand{\bi}{\begin{itemize}}
\newcommand{\ei}{\end{itemize}}
\newcommand{\bn}{\begin{enumerate}}
\newcommand{\en}{\end{enumerate}}
\newcommand{\bc}{\begin{center}}
\newcommand{\ec}{\end{center}}

\newcommand{\Dir}{\kern -6.4pt\Big{/}}
\newcommand{\Dirin}{\kern -10.4pt\Big{/}\kern 4.4pt}
\newcommand{\DDir}{\kern -8.0pt\Big{/}}
\newcommand{\DGir}{\kern -6.0pt\Big{/}}

\def\frac#1#2{ {{#1} \over {#2} }}

\def\beq{\begin{equation}}
\def\beeq{\begin{eqnarray}}
\def\eeq{\end{equation}}
\def\eeeq{\end{eqnarray}}
\def\a0{\bar\alpha_0}

\def\b0{\beta_0}

\def\ee{e^+e^-}

\def\lms{\Lambda^{(n_{\rm f}=4)}_{\overline{\mathrm{MS}}}}

\def\MSbar{\overline{\mathrm{MS}}}

\def\slashchar#1{\setbox0=\hbox{$#1$}           
     \dimen0=\wd0                                 
     \setbox1=\hbox{/} \dimen1=\wd1               
     \ifdim\dimen0>\dimen1                        
        \rlap{\hbox to \dimen0{\hfil/\hfil}}      
        #1                                        
     \else                                        
        \rlap{\hbox to \dimen1{\hfil$#1$\hfil}}   
        /                                         
     \fi}                                         %

\def\be{\begin{equation}}
\def\ee{\end{equation}}
\def\bea{\begin{eqnarray}}
\def\eea{\end{eqnarray}}

\def\slash{/\kern -5pt}

\def\ims #1 {\ensuremath{M2_{[#1]}}}

\def\s22w{s_{2W}^2}

\begin{document}

\begin{flushright}
{DFTT 05/2004}\\ 
{SHEP-03-28}\\ 
{February 2004}
\end{flushright}

\vspace*{2.0truecm}
\begin{center}
{\Large \bf
One-loop weak corrections to $\gamma/Z$ hadro-production\\[0.25cm]
at finite transverse momentum\footnote{Work supported in 
part by the U.K.\ Particle Physics and
Astronomy Research Council (PPARC),
by the European Union (EU) under contract HPRN-CT-2000-00149 and by the 
Italian Ministero dell'Istruzione, dell'Universit\`a e della Ricerca
(MIUR) under contract 2001023713\_006.}}
\\[1.5cm]
{\large Ezio Maina}\\[0.15 cm]
{\it Dipartimento di Fisica Teorica -- Universit\`a di Torino}\\
{\it and} \\
{\it Istituto Nazionale di Fisica Nucleare -- Sezione di Torino}\\
{\it Via Pietro Giuria 1, 10125 Torino, Italy}
\\[0.5cm]
{\large Stefano Moretti and Douglas A. Ross}\\[0.15 cm]
{\it School of Physics and Astronomy, University of Southampton}\\
{\it Highfield, Southampton SO17 1BJ, UK}\\[0.25cm]
\end{center}
\vspace*{2.0cm}
\begin{abstract}
We illustrate the effects of one-loop weak corrections onto the production
of neutral gauge bosons of the Standard Model at RHIC-Spin, Tevatron
and LHC, in presence of quark/gluon radiation from the initial state. We find 
such effects to be rather large, up to ${\cal O}(10-20\%)$ in typical
observables at all such colliders, where the
cross section is measurable, thus advocating their inclusion in 
precision analyses.
\end{abstract}
\vspace*{1.0cm}
\centerline{Keywords: Standard Model, Electroweak effects, Loop calculations, 
Hadron colliders}

\newpage

\section{Prompt-photon and on-shell $Z$ hadro-production}
\label{Sec:Intro}

The neutral-current processes ($V=\gamma,Z$)
\begin{equation}\label{procs_neutral}
q\bar q \to g V\quad{\rm{and}}\quad q(\bar q) g\to q(\bar q) V
\end{equation}
with $V\to\ell^+\ell^-$
are two of the cleanest probes of the partonic content of (anti)protons,
in particular of antiquark and gluon
densities. In order to measure the latter it is necessary to study
the vector boson $p_T$ spectrum. According to \cite{bego1,bego2} the gluon
density dominates for $p_T\ > Q/2$ where $Q$ is the lepton pair 
invariant mass.
In the presence of polarised beams these reactions give access to the
spin--dependent gluon distribution which is presently only poorly known.
Thanks to the introduction of improved algorithms 
\cite{Frixione:1999pl}--\cite{Frixione:1999ya} for the 
selection of (prompt) photons generated in the hard scatterings  
(\ref{procs_neutral}), 
as opposed to those generated in the fragmentation of the accompanying
gluon/quark jet, and to the high experimental resolution achievable
in reconstructing 
$Z\to\ell^+\ell^-$ ($\ell=e,\mu$) decays, they are regarded -- together
with the twin charged-current channels 
\begin{equation}\label{procs_charged}
q\bar q' \to g W \quad{\rm{and}}\quad q(\bar q) g\to q'(\bar q')W,
\end{equation}
wherein $W\to\ell\nu_\ell$ -- as precision observables in hadronic 
physics. In fact, in some instances, accuracies
of order one percent are expected to be 
attained in measuring these processes \cite{reviews},
both at present and future proton-(anti)proton 
experiments. These include the Relativistic Heavy Ion Collider running
with polarised proton beams (RHIC-Spin) at BNL ($\sqrt s_{pp}=300-600$ GeV), 
the Tevatron collider at FNAL (Run 2, $\sqrt s_{p\bar p}=2$ TeV) 
and the Large Hadron Collider (LHC) at CERN
($\sqrt s_{pp}=14$ TeV).

Not surprisingly then, a lot of effort has been spent over the years
in computing higher order corrections to all such Drell--Yan type processes.
To stay with the
neutral-current ones, the subject of this paper, these include 
next-to-leading order (NLO) QCD
calculations of both prompt-photon \cite{coka,govo} and vector boson
production \cite{kamal}. QCD corrections to the $p_T$ distributions have been
computed in Refs.~\cite{elma,arre}.
As for the full $\cal O (\alpha)$ Electro-Weak (EW)
corrections to $Z$ production and continuum
neutral-current processes (at zero transverse momentum), these have been 
completed in \cite{Baur:2001ze} (see also \cite{Haywood:1999qg}), building on
the calculation of the QED part in \cite{Baur:1998zf}.

In the case of polarised (anti)proton beams, the process of calculating 
higher order corrections has proceeded more slowly \cite{weber,gehr}.
NLO QCD corrections to the transverse momentum spectrum of Drell-Yan type 
processes via neutral-currents
in presence of (longitudinal) spin effects from the initial state can be found
for the non--singlet case in \cite{chco1,chco2}, while the complete
calculation  has been recently published in Ref.~\cite{Ravindran:2002na}
(see also Ref.~\cite{vanNeerven:2002sk}).

The relatively large impact of one-loop EW corrections, as
compared to the QCD ones, can be understood 
(see Refs.~\cite{Melles:2001ye}--\cite{Denner:2001mn} and references therein
for reviews) in terms of the so-called
Sudakov (leading) logarithms of the form 
$\alpha_{\mathrm{W}}\log^2(\sqrt{\hat{s}}/M_{W}^2)$, which appear
in the presence of higher order weak corrections
(hereafter, $\alpha_{\rm{W}}\equiv\alpha_{\rm{\small EM}}/\sin2\theta_{\rm W}$,
with $\alpha_{\rm{\small EM}}$ the Electro-Magnetic (EM) coupling constant and
$\theta_{\rm W}$ the weak mixing angle)\footnote{In some cases, 
        leading ($\sim\alpha_{\mathrm{W}}^n\log^{2n  }(s/M_W2)$),
    sub-leading ($\sim\alpha_{\mathrm{W}}^n\log^{2n-1}(s/M_W2)$) and
sub-sub-leading ($\sim\alpha_{\mathrm{W}}^n\log^{2n-2}(s/M_W2)$) can be 
resummed.}.
These `double logs' are due to a lack of cancellation of infrared (both soft
and collinear) virtual and real emission in
higher order contributions due to $W$-exchange
in spontaneously broken non-Abelian theories.

The problem is, in principle, present also in QCD. In practice, however, 
it has no observable consequences, because of the averaging on the 
colour degrees of freedom of partons, forced by their confinement
into colourless hadrons. This does not occur in the EW case,
where, e.g., the initial state can have a non-Abelian charge,
dictated by the given collider beam configuration. Modulo the
effects of the Parton Distribution Functions (PDFs), which spoil the subtle
cancellations among subprocesses with opposite non-Abelian charge,
 for example, this argument holds
for an initial quark doublet in proton-(anti)proton scatterings. These
logarithmic corrections (unless the EW process is mass-suppressed)
are universal (i.e., process independent) and are finite (unlike in
QCD), as the masses of the EW gauge bosons provide a physical
cut-off for $W$-boson emission. Hence, for typical experimental
resolutions, softly and collinearly emitted weak bosons need not be included
in the production cross-section and one can restrict oneself to the calculation
of weak effects originating from virtual corrections. In fact, one should 
recall that real weak bosons are unstable and decay into high
transverse momentum leptons and/or jets, which are normally
captured by the detectors. In the definition of an
exclusive cross section then,
one tends to remove events with such additional particles.
Under such circumstances,
the (virtual) exchange of $Z$-bosons also generates similar logarithmic
corrections, 
$\alpha_{\mathrm{W}}\log^2(\sqrt{\hat{s}}/M_{Z}^2)$.
Besides, the genuinely weak contributions can  be
isolated in a gauge-invariant manner from purely EM effects,
at least in some simpler cases -- which do include processes 
(\ref{procs_neutral}) but not (\ref{procs_charged})  -- and the latter may or may not
be included in the calculation, depending on the observable being studied. 

A further aspect that should be recalled is that weak corrections naturally
introduce parity-violating effects in observables, detectable through
asymmetries in the cross-section, which are often regarded as an indication
of physics beyond the Standard Model
(SM) \cite{reviews,Maina:2003is,Dittmar:2003ir}. 
These effects are further enhanced if polarisation
of the incoming beams is exploited, such as at RHIC-Spin
\cite{Bourrely:1990pz,Ellis:2001ba}.
Comparison of theoretical predictions 
involving parity-violation with experimental data 
is thus used as another powerful tool for confirming or 
disproving  the existence of some beyond the SM scenarios, such as those 
involving right-handed weak currents \cite{Taxil:1997kj}, contact interactions
\cite{Taxil:1996vf} and/or new massive gauge bosons 
\cite{Taxil:1998ni,Taxil:1996vs}.

In view of all this,  it becomes of crucial importance to assess
the quantitative relevance of weak corrections
affecting processes (\ref{procs_neutral})--(\ref{procs_charged}).
It is the aim of our paper to report on the computation of the full
one-loop weak effects entering processes (\ref{procs_neutral})
while the study of those for (\ref{procs_charged}) will be deferred
to a future publication \cite{preparation}.

\section{Calculation and results}
\label{SubSec:Results}

\begin{figure}[!h]
\begin{center}
\begin{picture}(365,225)
\SetScale{1.0}
\Photon(0,175)(50,175){4}{4}
\ArrowLine(150,125)(125,137) \Line(125,137)(100,150)
\ArrowLine(100,150)(75,162) \ArrowLine(75,162)(50,175)
\ArrowLine(50,175)(150,225) \Gluon(115,142)(150,150){4}{3}
\PhotonArc(87,155)(12,337,150){3}{5} \put(87,175){$W,Z$}
\put(10,185){$\gamma,Z$}

\Photon(200,175)(250,175){4}{4}
\ArrowLine(350,125)(325,137) \ArrowLine(325,137)(300,150)
\ArrowLine(300,150)(275,162) \ArrowLine(275,162)(250,175)
\ArrowLine(250,175)(350,225) \Gluon(275,162)(350,175){4}{8}
\PhotonArc(312,143)(12,334,150){3}{5} \put(324,152){$W,Z$}
\put(210,185){$\gamma,Z$}

\Photon(0,50)(50,50){4}{4}
\ArrowLine(150,0)(100,25) \ArrowLine(100,25)(50,50)
\ArrowLine(50,50)(90,72) \ArrowLine(90,72)(110,82)
\ArrowLine(110,82)(150,100) \Gluon(100,25)(150,35){4}{5}
\PhotonArc(100,76)(12,32,206){3}{5} \put(87,95){$W,Z$}
\put(10,60){$\gamma,Z$}

\Photon(200,50)(240,50){4}{4} \Photon(265,50)(300,50){4}{4}
\ArrowLine(350,0)(325,25) \ArrowLine(325,25)(300,50)
\ArrowLine(300,50)(350,100) \Gluon(321,29)(350,45){4}{3}
\GCirc(252,50){12}{.5}
\put(210,60){$\gamma,Z$}
\end{picture}

\begin{picture}(365,225)

\Photon(100,175)(150,175){4}{4}
\ArrowLine(250,125)(225,137) \ArrowLine(225,137)(200,150)
\ArrowLine(200,150)(175,162) \ArrowLine(175,162)(150,175)
\ArrowLine(150,175)(250,225) \Gluon(200,150)(250,170){4}{5}
\PhotonArc(200,150)(24,154,330){3}{8} \put(187,115){$W,Z$}
\put(110,185){$\gamma,Z$}

\Photon(0,50)(50,50){4}{4}
\ArrowLine(150,0)(125,12) \ArrowLine(125,12)(87,30)
 \ArrowLine(87,30)(50,50)
\ArrowLine(50,50)(87,70) \ArrowLine(87,70)(150,100)
 \Gluon(115,17)(150,25){4}{3}
\Photon(97,25)(97,75){3}{5} \put(102,50){$W,Z$}
\put(10,60){$\gamma,Z$}

\Photon(200,50)(250,50){4}{4}
\ArrowLine(350,0)(325,12) \ArrowLine(325,12)(297,25)
 \Photon(297,25)(250,50){3}{5}
\Photon(250,50)(297,75){-3}{5} \ArrowLine(297,75)(350,100)
 \Gluon(315,17)(350,25){4}{3}
\ArrowLine(297,25)(297,75) \put(275,73){$W$} \put(275,19){$W$}
\put(210,60){$\gamma,Z$}

\end{picture}

\begin{picture}(365,105)

\Photon(0,50)(50,50){4}{4}
\put(10,60){$\gamma,Z$}
\ArrowLine(150,29)(116,38) \ArrowLine(116,38)(83,44) \ArrowLine(83,44)(50,50)
\ArrowLine(50,50)(116,83) \ArrowLine(116,83)(150,100)
\Gluon(83,44)(150,0){-4}{8} \Photon(116,38)(116,83){4}{5}
\put(122,55){$W,Z$}

\Photon(200,50)(250,50){4}{4}
\ArrowLine(350,0)(307,20) 
 \Photon(307,20)(250,50){3}{5}
\Photon(250,50)(307,80){-3}{5} \ArrowLine(307,80)(350,100)
 \Gluon(307,50)(350,50){4}{5}
\ArrowLine(307,20)(307,50) \ArrowLine(307,50)(307,80)
 \put(275,73){$W$} \put(275,19){$W$}
\put(210,60){$\gamma,Z$}

\end{picture}
\end{center}
\caption{\small Graphs describing processes (\ref{procs_neutral}) in the presence
of one-loop weak corrections. The shaded blob
represents all the contributions to the gauge boson 
self-energy and is dependent on the Higgs mass (we have set
$M_H=115$ GeV). We neglect loops involving the
Higgs boson coupling to the fermion line. The graphs in which the 
exchanged gauge boson is a $W$-boson
are accompanied by those in which the latter is replaced
by its corresponding Goldstone boson. There is a similar set of diagrams
in which the direction of the fermion line is reversed, with the exception
of the last graph, as here reversal does {not} lead to a distinct topology.}
\label{fig:graphs}
\end{figure}
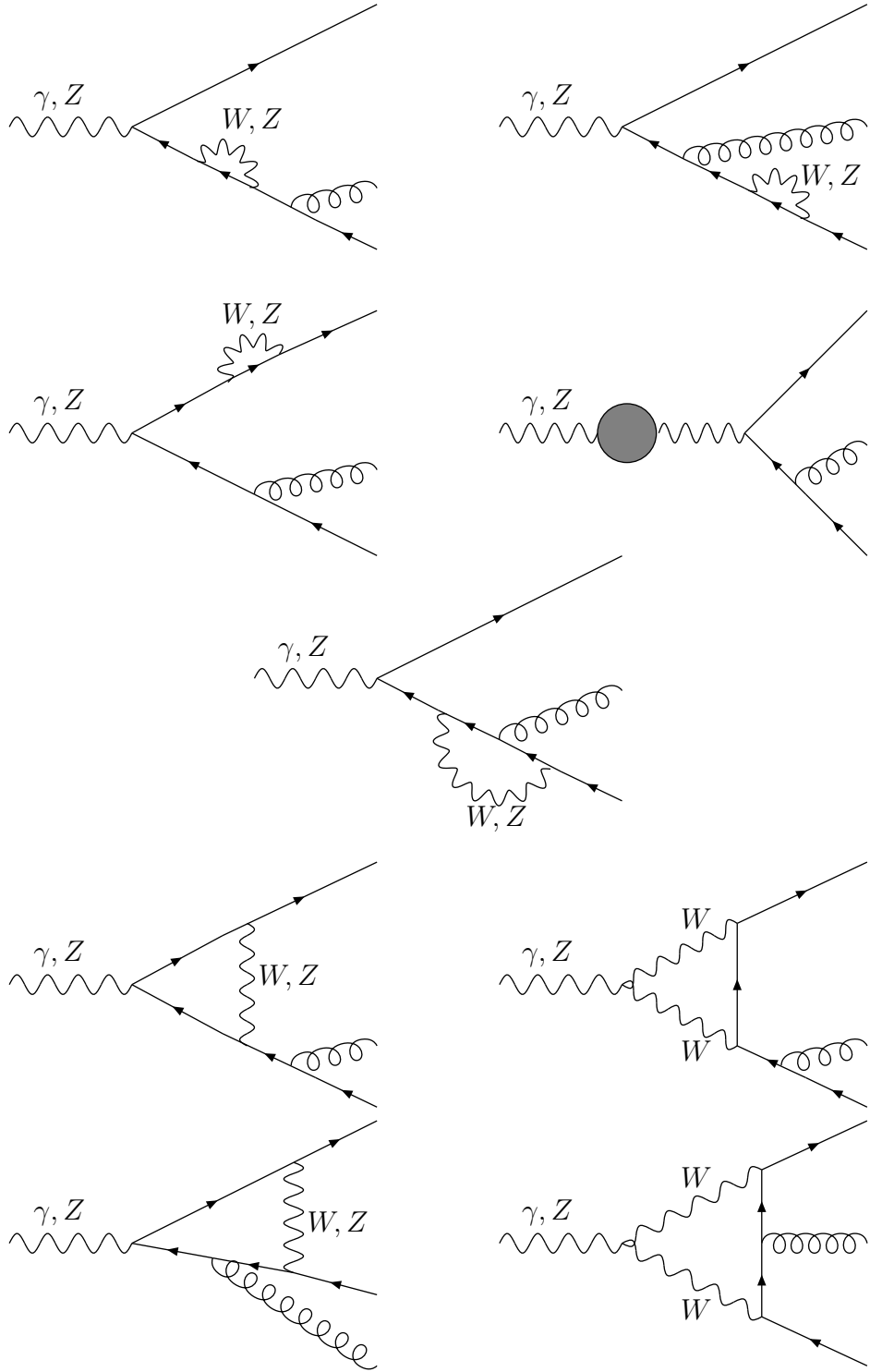

Since we are considering weak corrections that may be
identified via their induced parity-violating effects and since we wish to
apply our results to the case of polarised proton beams, it is convenient to 
work in terms of helicity Matrix Elements
(MEs). Here, we define the helicity amplitudes by using the formalism
discussed in Ref.~\cite{Maina:2002wz}. 
At one-loop level such helicity amplitudes   
acquire higher order corrections from: 
  (i) self-energy insertions on the fermions and gauge bosons;
 (ii) vertex corrections and
(iii) box diagrams. All such contributions
are pictured in Fig.~\ref{fig:graphs}. 

The self-energy and vertex correction graphs contain ultraviolet divergences
that have been subtracted here by using the `modified' Minimal Subtraction
($\MSbar$) scheme at
the scale $\mu=M_Z$. Thus the couplings are taken to be
those relevant for such a subtraction: e.g., the EM coupling,
$\alpha_{\mathrm{EM}}$, has been taken to be $1/128$ at the above subtraction
point. The one exception to this
renormalisation scheme has been the case of the self-energy insertions
on external fermion lines, which have been subtracted on mass-shell,
so that the external fermion fields create or destroy particle states
with the correct normalisation.

All these graphs are infrared and collinear convergent so that they
 may be expressed in terms of Passarino-Veltman \cite{VP} functions
which are then evaluated numerically. The expressions for
 each of these diagrams 
have been calculated using FORM \cite{FORM} and checked by an
independent program based on FeynCalc \cite{FeynCalc}. For the numerical
evaluation of the scalar integrals we have relied on the FORTRAN
package FF \cite{FF1.9}. 
A further check on our results has been carried out
by setting the polarisation vector of the $V$-boson 
proportional to its momentum
and verifying that the sum of all one-loop diagrams
vanishes, as required by gauge and BRST invariance.
The full expressions for the contributions from these graphs are too
lengthy to be reproduced here. 

In both processes in (\ref{procs_neutral}), external
(anti)quarks have been taken 
massless and both vector bosons ($V=\gamma,Z$) have been put  
on-shell. In contrast, the top quark entering the loops in both reactions has
been assumed to have the mass $m_t=175$ GeV. The $Z$ mass used was
$M_Z=91.19$ GeV and was related to the $W$-mass, $M_W$, via the
SM formula $M_W=M_Z\cos\theta_W$, where $\sin2\theta_W=0.232$.
(Corresponding widths were $\Gamma_Z=2.5$ GeV and $\Gamma_W=2.08$ GeV.)
For the strong coupling constant, $\alpha_{\rm S}$, we have used the 
one-loop expression at $Q^2=\sqrt{\hat s}$ with $\lms$ chosen to match 
the value required by the (LO) PDFs used. The latter were Gehrmann-Stirling 
set A for RHIC and Martin-Roberts-Stirling-Thorne set 2001 LO
for Tevatron/LHC  \cite{PDFs}.

\subsection{RHIC}

\begin{figure}[!t]
\begin{center}
\vspace{-1.cm}
{\epsfig{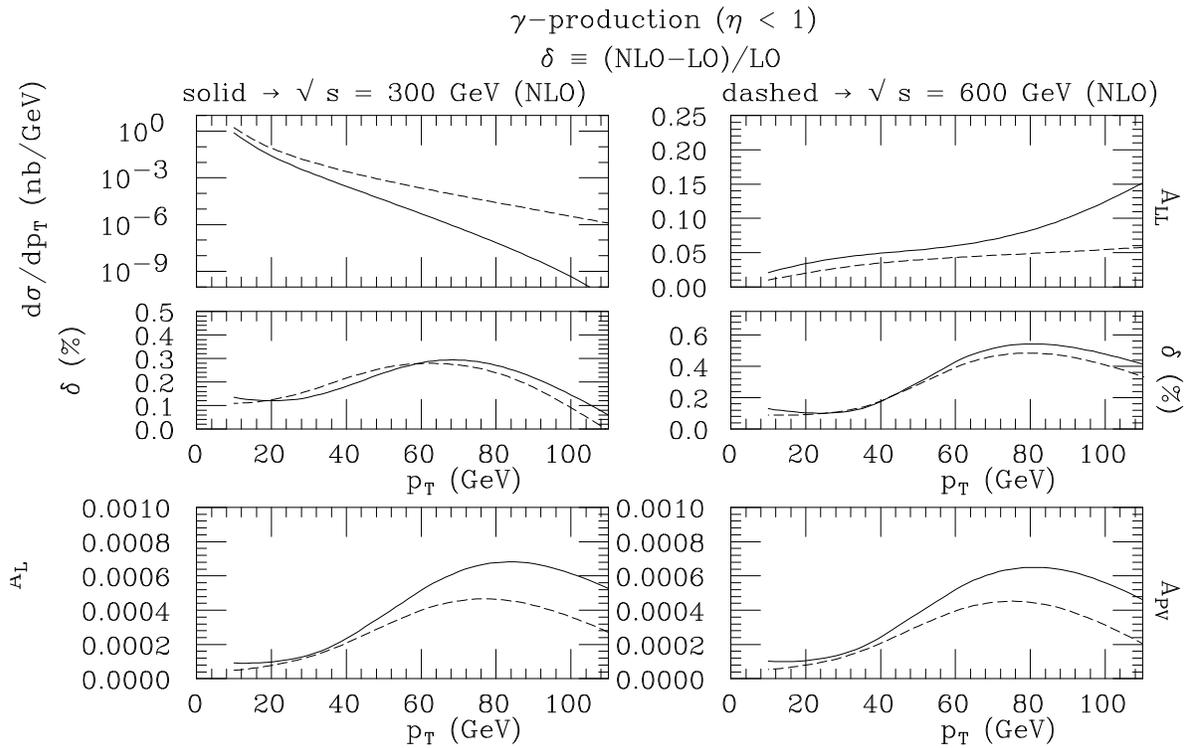}}
\end{center}
\vspace*{-0.75cm}
\caption{\small The transverse momentum dependence of the $\gamma$-boson cross 
section in (\ref{procs_neutral}) as well as of the beam asymmetries
in (\ref{asymmetries}) at NLO
(large frames) and the size of the one-loop weak 
corrections (small frames, limitedly to case in which
the latter is non-zero), at RHIC-Spin ($\sqrt s_{pp}=300$ 
and 600 GeV). Notice that the pseudorapidity range of either
particle in the final state is limited to $|\eta|<1$.}
\label{fig:pTG-RHIC}
\end{figure}

\begin{figure}[!t]
\begin{center}
\vspace{-1.cm}
{\epsfig{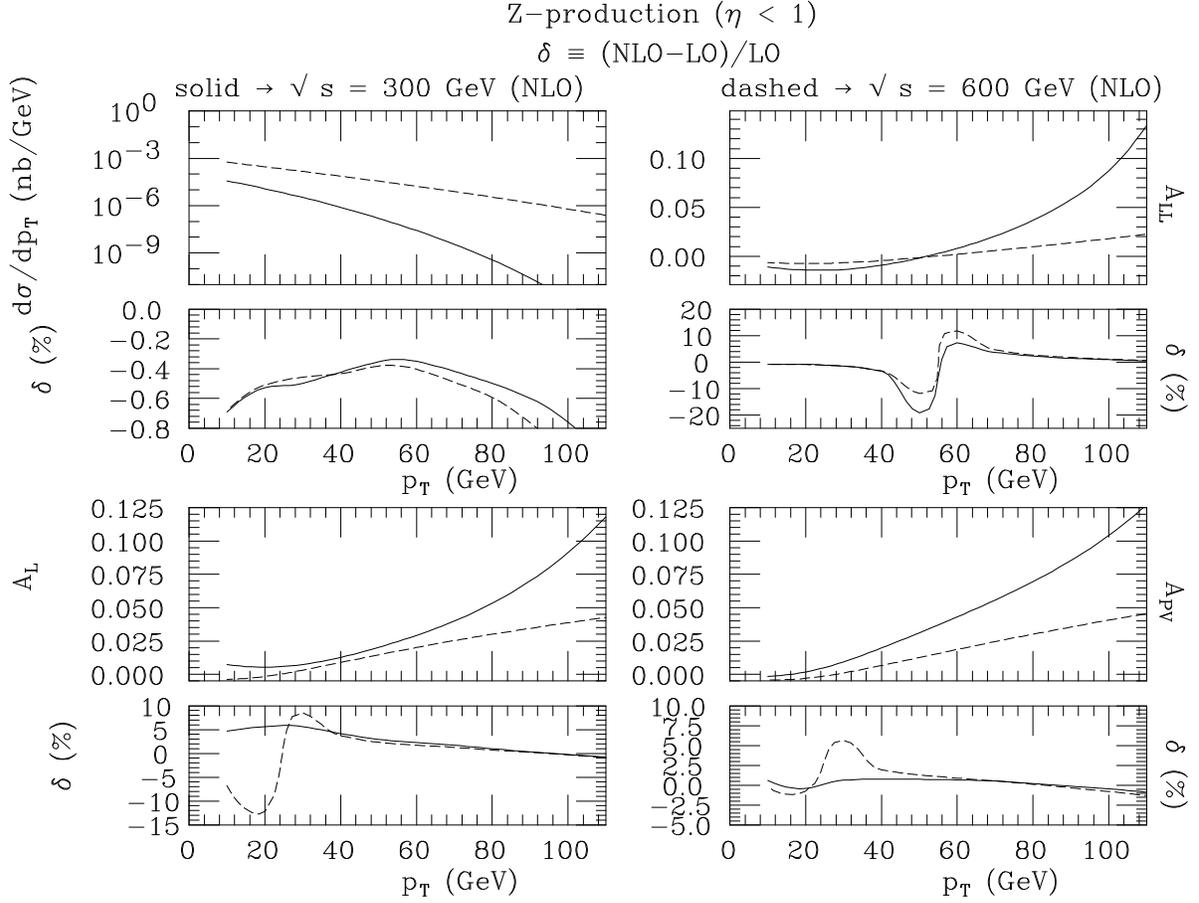}}
\end{center}
\vspace*{-0.75cm}
\caption{\small The transverse momentum dependence of the $Z$-boson cross 
section in (\ref{procs_neutral}) as well as of the beam asymmetries
in (\ref{asymmetries}) at NLO
(large frames) and the size of the one-loop weak 
corrections (small frames), at RHIC-Spin ($\sqrt s_{pp}=300$ 
and 600 GeV). Notice that the pseudorapidity range of either
particle in the final state is limited to $|\eta|<1$.}
\label{fig:pTZ-RHIC}
\end{figure}

The following beam asymmetries can, e.g., be defined at RHIC-Spin:
\begin{eqnarray}\label{asymmetries}                                \nonumber
A_{LL} \, d\sigma &\equiv &\,d\sigma_{++}\,   - \, d\sigma_{+-} \\ \nonumber 
                  &+      &\,d\sigma_{--}\,   - \, d\sigma_{-+},\\ \nonumber
~~A_L \,  d\sigma &\equiv &\,d\sigma_{- }\, ~~- \, d\sigma_{+ },\\ 
A_{PV} \, d\sigma &\equiv &\,d\sigma_{--}\,   - \, d\sigma_{++}.
\end{eqnarray}
The first two are parity-conserving while the last two are parity-violating.
Figs.~\ref{fig:pTG-RHIC}--\ref{fig:pTZ-RHIC} show the NLO distributions in such quantities
for both $\gamma$ and $Z$ final states, alongside those for the total cross
section, as a function of the transverse momentum $p_T$, within the pseudo-rapidity
range $|y|<1$. The corrections due to full one-loop weak effects are also presented 
(for the cases in which the Born level result is non-zero). Effects onto the total
cross sections are rather small, below the percent level, as expected, because
of the rather low centre-of-mass
(CM) energy available at partonic level, which is comparable with $M_W$
and $M_Z$, so that logarithmic corrections are not enhanced. Nonetheless, in the case
of $\gamma$-production, at $p_T=10$ GeV, the total NLO yearly rates of approximately
150,000 and 1,350,000 events accessible at low and high energy, respectively,
contain a sizable contribution due to purely weak effects, about 200 and 1,500
events in correspondence of $\sqrt s_{pp}=300$ and 600 GeV (for the values of
luminosity 200 and 800 pb$^{-1}$, respectively). In case 
of $Z$-production, only at 600 GeV NLO effects are sizable,
as they are responsible for 3 events being subtracted (the correction is negative)
to the LO prediction of 457 events (at $p_T=10$ GeV). (The LO rate at
300 GeV for $Z$-production at such a transverse momentum is of only 7 events,
unaffected by NLO weak effects.)

Rather large effects
do appear in general for the asymmetries, particularly for the case of $Z$ boson
final states. In fact, for the latter, in the case of $A_L$ and $A_{LL}$, they can range
up to $\pm(15-20)\%$, while they are somewhat lower for the case of $A_{PV}$, 5\% or so.
Unfortunately, none of such NLO effects on the asymmetries
is detectable, because of the poor
production rate of $Z$-bosons. 
For photonic final states, one-loop weak effects on $A_{LL}$ are not
much larger than those on the total rates, nonetheless, they might just be
observable at low $p_T$. For the cases of
$A_L$ and $A_{PV}$, which for the photon
are exactly zero at Born level in massless QCD, one-loop
weak effects are too poor to be observed experimentally.

\subsection{Tevatron and LHC}

\begin{figure}[!h]
\vspace*{2.75cm}
\begin{center}
{\epsfig{file=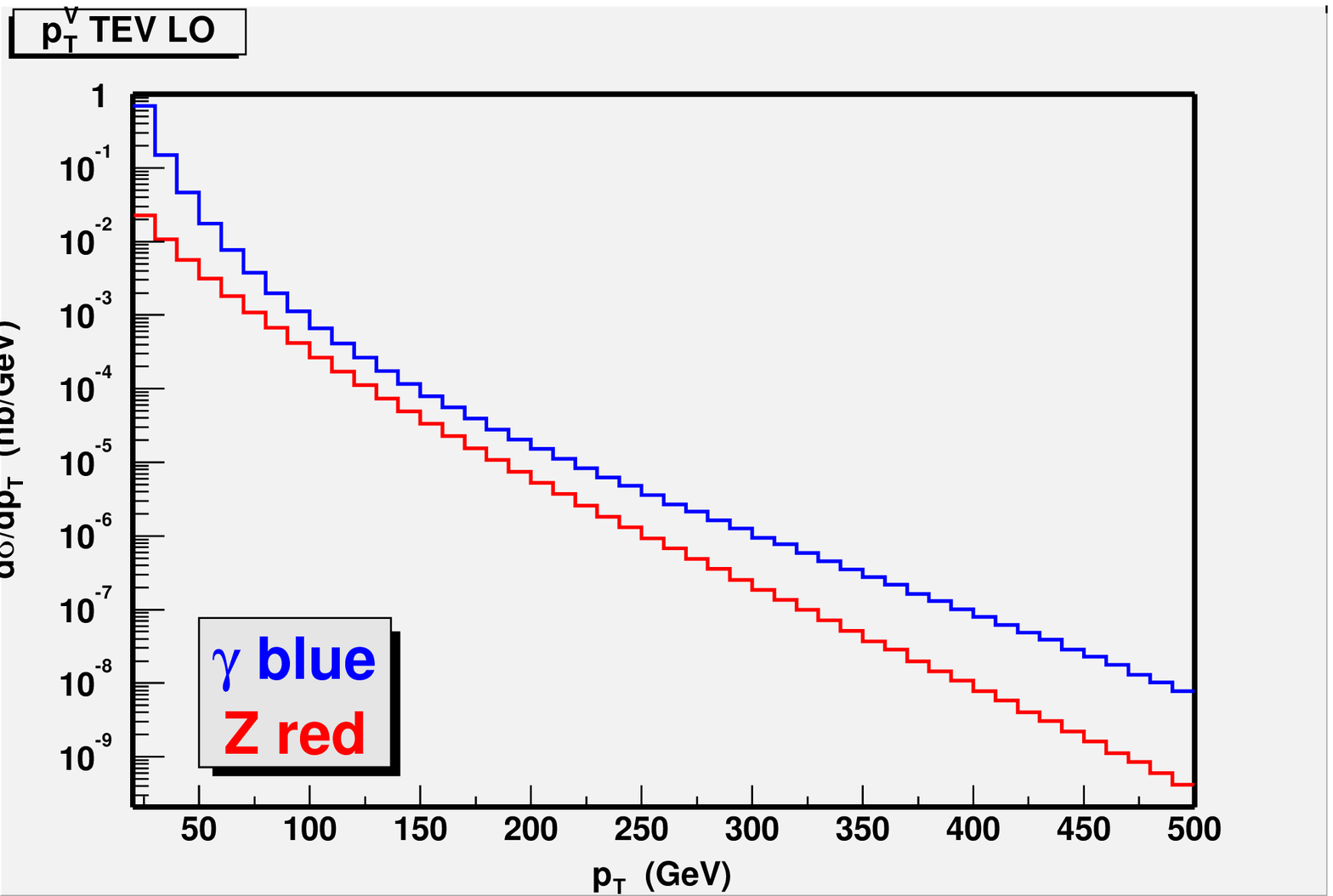,  width=10cm, angle=0}}
{\epsfig{file=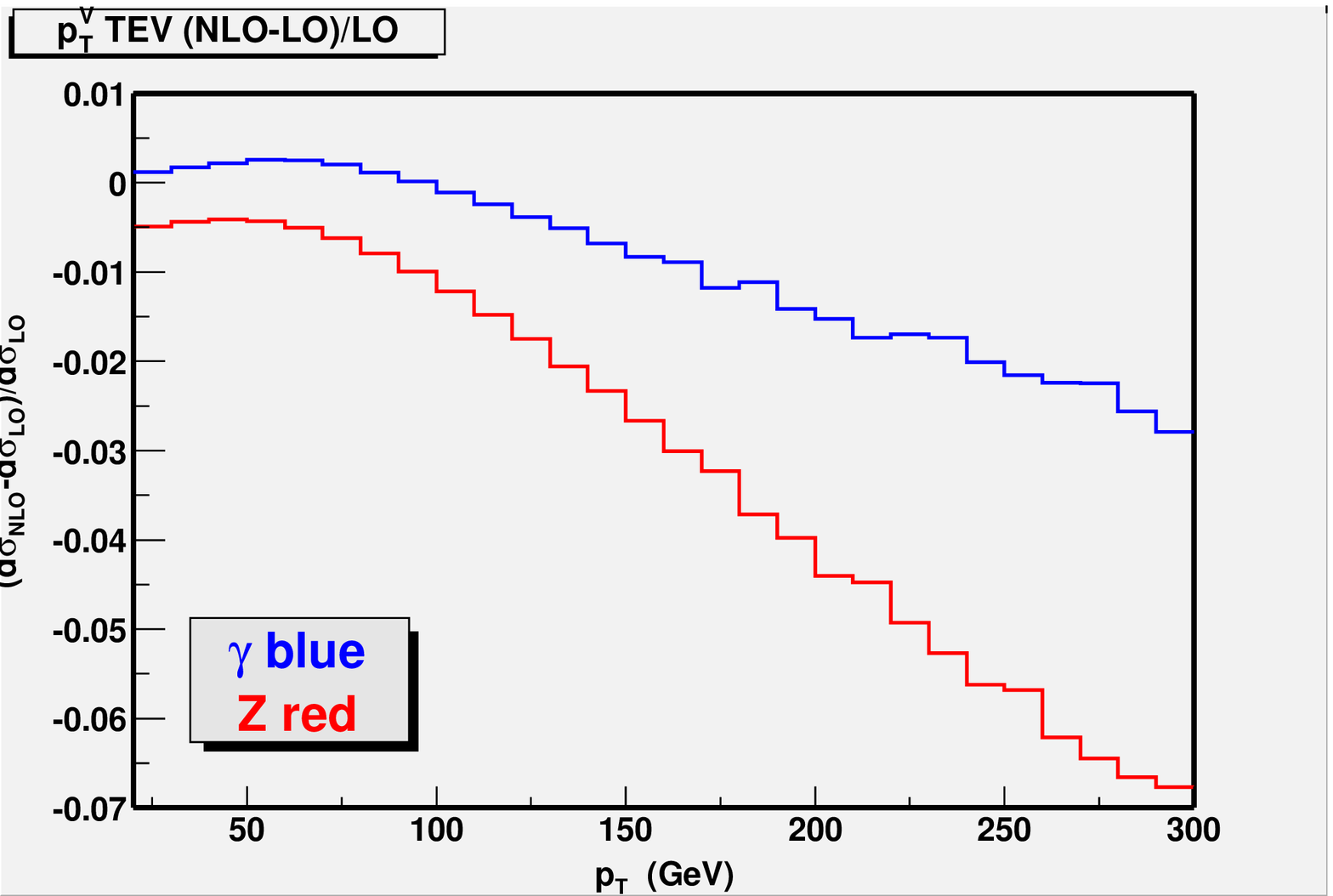, width=10cm, angle=0}}
\end{center}
\vspace*{-0.75cm}
\caption{\small The transverse momentum dependence of the $\gamma$- and
$Z$-boson cross 
sections in (\ref{procs_neutral}) at LO 
(top frame) and the size of the one-loop weak 
corrections (bottom frame), at Tevatron ($\sqrt s_{p\bar p}=2$ TeV).
Notice that the pseudorapidity range of the jet
in the final state is limited to $|\eta|<3$.} 
\label{fig:pT-Tev}
\end{figure}

\begin{figure}[!h]
\vspace*{2.75cm}
\begin{center}
{\epsfig{file=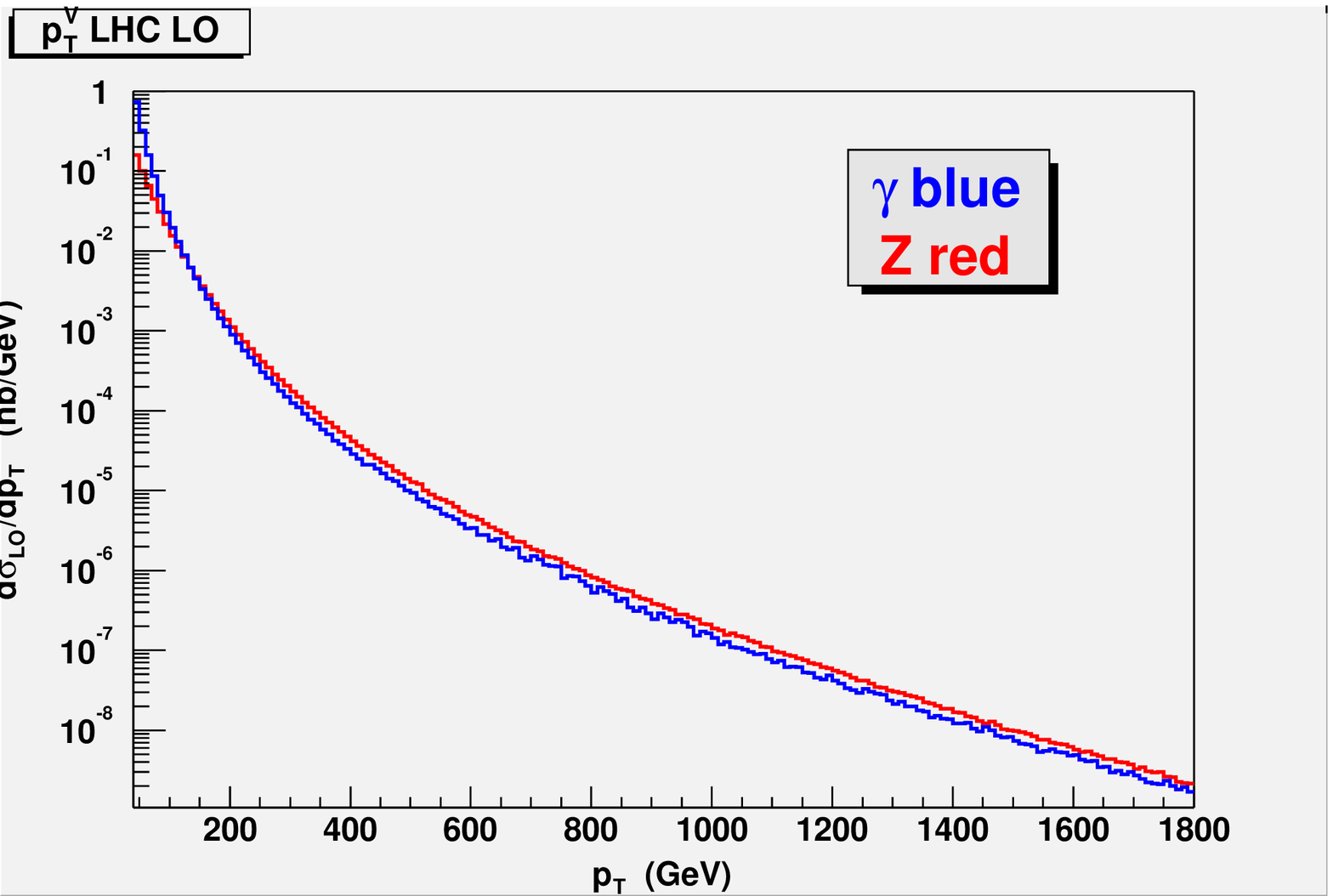,  width=10cm, angle=0}}
{\epsfig{file=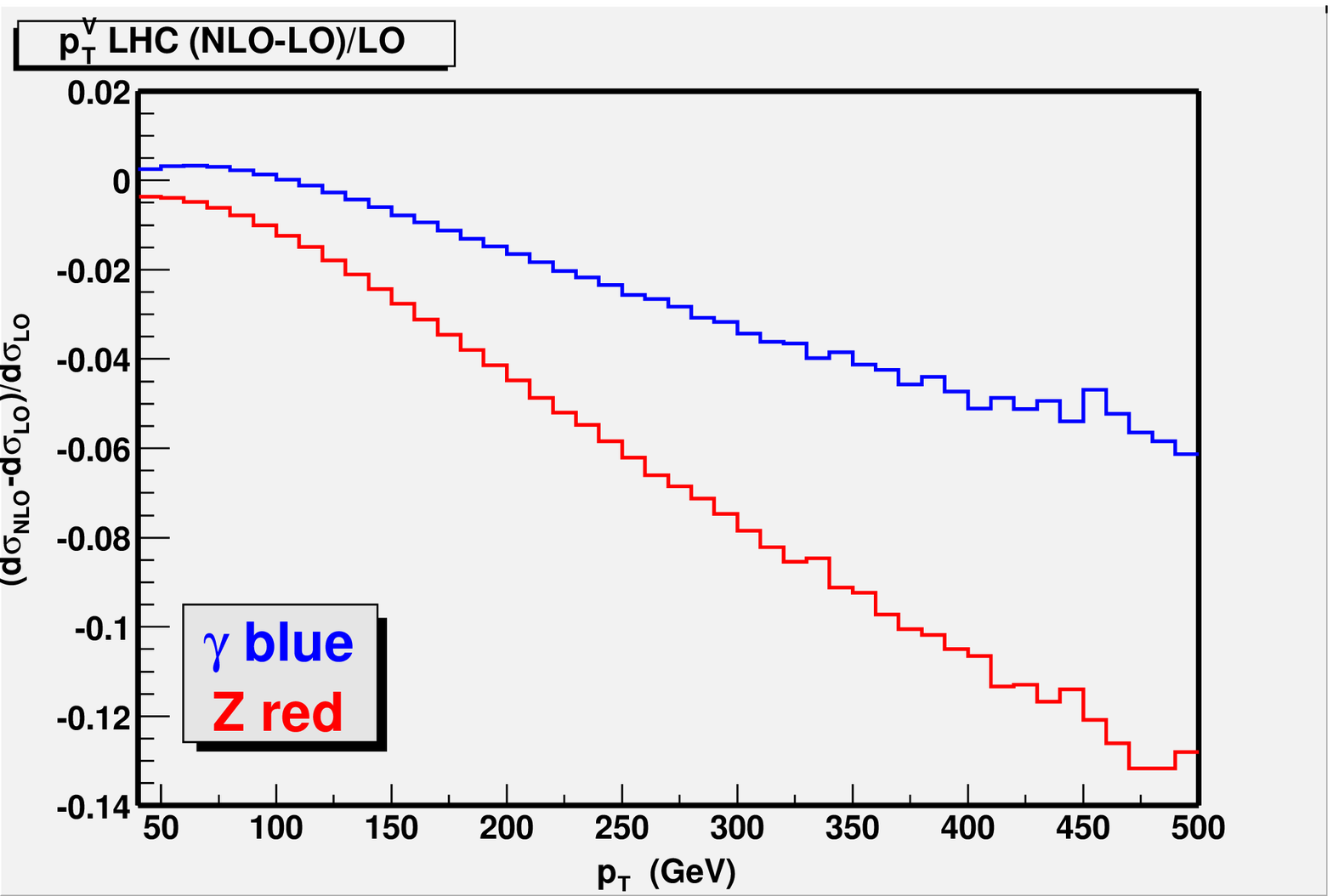, width=10cm, angle=0}}
\end{center}
\vspace*{-0.75cm}
\caption{\small The transverse momentum dependence of the $\gamma$- and
$Z$-boson cross 
sections in (\ref{procs_neutral}) at LO 
(top frame) and the size of the one-loop weak 
corrections (bottom frame), at LHC ($\sqrt s_{pp}=14$ TeV).
Notice that the pseudorapidity range of the jet
in the final state is limited to $|\eta|<4.5$.} 
\label{fig:pT-LHC}
\end{figure}

Figs.~\ref{fig:pT-Tev}--\ref{fig:pT-LHC} show the effects of the 
${\cal O}(\alpha_{\rm{S}}\alpha_{\rm{EW}}^2)$
terms relatively to the ${\cal O}(\alpha_{\rm{S}}\alpha_{\rm{EW}})$
Born results ($\alpha_{\rm{EM}}$ replaces $\alpha_{\rm{EW}}$ for photons),
as well as the absolute magnitude of the latter, as a function
of the transverse momentum, at Tevatron and LHC, respectively. 
The corrections are found to be rather large at both colliders, particularly
for $Z$-production. In case of the latter,
such effects are of order --7\% at Tevatron for $p_T\approx 300$ GeV
and --14\% at LHC for $p_T\approx 500$ GeV. In general, above 
$p_T\approx100$ GeV,
they tend to (negatively) increase, more or less linearly, with $p_T$.
Such effects will be hard to observe at Tevatron but
will indeed be observable at LHC. 
For example, at FNAL, for $Z$-production and decay into electrons and muons
with BR$(Z\rightarrow e,\mu)\approx 6.5\%$, assuming
$L= 2-20$ fb$^{-1}$ as integrated luminosity, in
a window of 10 GeV at $p_T = 100$ GeV, one finds
500--5000 $Z+j$ events at LO, hence a
$\delta\sigma/\sigma\approx -1.2\%$ EW NLO correction corresponds to only
6--60 fewer 
events. At CERN, for the same production and decay channel, assuming now 
$L= 30$ fb$^{-1}$, in a window of 40 GeV at $p_T = 450$ GeV, 
we expect about 2000 $Z+j$ events from LO, so that a 
$\delta\sigma/\sigma\approx -12\%$ EW NLO correction 
corresponds to 240 fewer events. In line with the normalisations seen
in the top frames of  Figs.~\ref{fig:pT-Tev}--\ref{fig:pT-LHC}
and the size of the corrections
in the bottom ones, absolute rates for the photon are similar
to those for the massive gauge boson while ${\cal O}(\alpha_{\rm{S}}\alpha_{\rm{EW}}^2)$ corrections are about a factor of two 
smaller.

\section{Conclusions}
\label{Sec:Summa}

Altogether, the results presented here point to the relevance of one-loop
${\cal O}(\alpha_{\rm{S}}\alpha_{\rm{W}}^2)$ weak contributions for precision
analyses of prompt-photon and neutral Drell-Yan events at both Tevatron and
LHC, also
recalling that the residual scale dependence of the known
higher order QCD corrections
to processes of the type (\ref{procs_neutral}) is very small 
in comparison \cite{arre}. Another relevant aspect is
that such higher order weak terms introduce parity-violating
effects in hadronic observables \cite{Ellis:2001ba}, which might just 
be observable at (polarised) RHIC-Spin, particularly in the case of
photons. (The case for polarised beams at the LHC, enabling the study
of parity-violating asymmetries on the same footing as at RHIC-Spin,
is currently being discussed as one of the possible upgrades of the 
CERN collider.) The size of the mentioned
corrections, relative to the lowest order results, 
is insensitive to the choice of PDFs. EM effects were neglected here
because they are not subject to logarithmic enhancement, thus
smaller with respect to the weak ones, or parity-violating effects either. 
However, their computation
is currently in progress \cite{preparation}. 

\section*{Acknowledgements}

SM and DAR are grateful to the Theoretical Physics department in Torino
for hospitality while part of this work was been carried out. Similarly,
EM thanks the High Energy Physics group at Southampton for arranging his
visit. This research is supported in part
by a Royal Society Joint Project within
the European Science Exchange Programme (Grant No. IES-14468).

\newpage
\thispagestyle{empty}

\newpage

\clearpage

\clearpage

\clearpage

\clearpage
\include{ErratumNewSingleBoson}

\end{document}

%% file: ErratumNewSingleBoson.tex
\begin{center}
{\Large \bf
Erratum to:\\
 One-loop weak corrections to $\gamma/Z$ hadro-production\\[0.25cm]
at finite transverse momentum\\[0.25cm]
[Phys. Lett. {\bf B593} (2004) 143-150]}
\\[1.5cm]
{\large Ezio Maina}\\[0.15 cm]
{\it Dipartimento di Fisica Teorica -- Universit\`a di Torino}\\
{\it and} \\
{\it Istituto Nazionale di Fisica Nucleare -- Sezione di Torino}\\
{\it Via Pietro Giuria 1, 10125 Torino, Italy}
\\[0.5cm]
{\large Stefano Moretti and Douglas A. Ross}\\[0.15 cm]
{\it School of Physics and Astronomy, University of Southampton}\\
{\it Highfield, Southampton SO17 1BJ, UK}\\[0.25cm]
\end{center}
\vspace*{1.0cm}
The results presented in Fig. 4 of the original paper 
mistakenly refer to a $pp$ collider 
of $\sqrt s_{pp}=2$ TeV instead of a $p \bar p$ one.
The correct results for the  
effects of the  ${\cal O}(\alpha_{\rm{S}}\alpha_{\rm{EW}}^2)$
terms relatively to the ${\cal O}(\alpha_{\rm{S}}\alpha_{\rm{EW}})$
Born results ($\alpha_{\rm{EM}}$ replaces $\alpha_{\rm{EW}}$ for photons),
as well as the absolute magnitude of the latter, as a function
of the transverse momentum at Tevatron are shown in Fig.~\ref{fig:pT-Tev}
below. 
The corrections are of order --6\%  for $Z+j$ production at
Tevatron for $p_T\approx 300$ GeV.
Since both the size of the corrections and the cross section for moderate values
of $p_T$ are similar to those for a $pp$ collider, our conclusions that 
such effects will be hard to observe at Tevatron but
will indeed be observable at LHC are unchanged.

\section*{Acknowledgements}
We are grateful to J.H.~K\"uhn, A.~Kulesza, S.~Pozzorini, M.~Schulze
\cite{KKPS} for pointing out the discrepancy with their calculation.

\begin{figure}[!h]
\begin{center}
{\epsfig{file=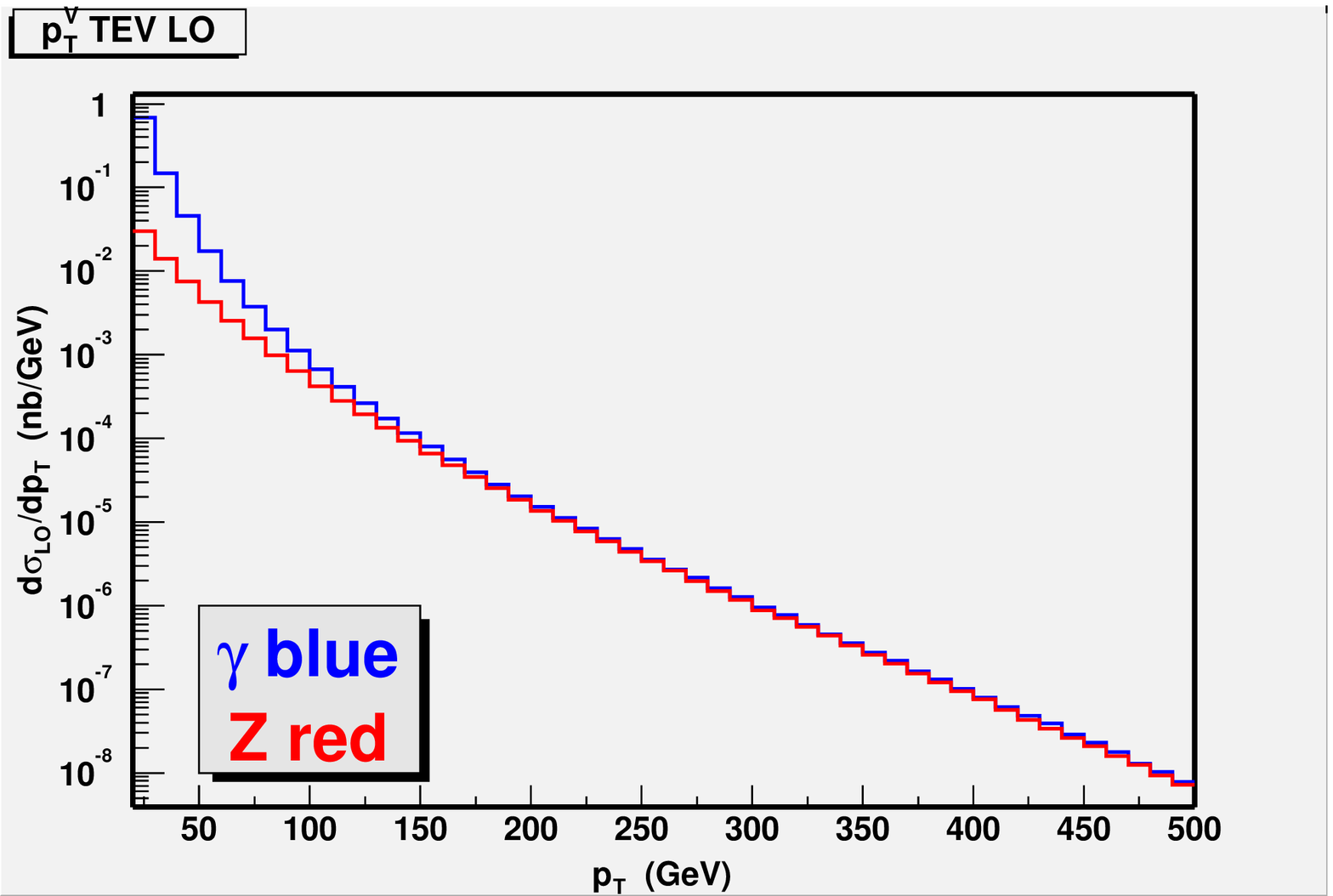,  width=15cm, angle=0}}
{\epsfig{file=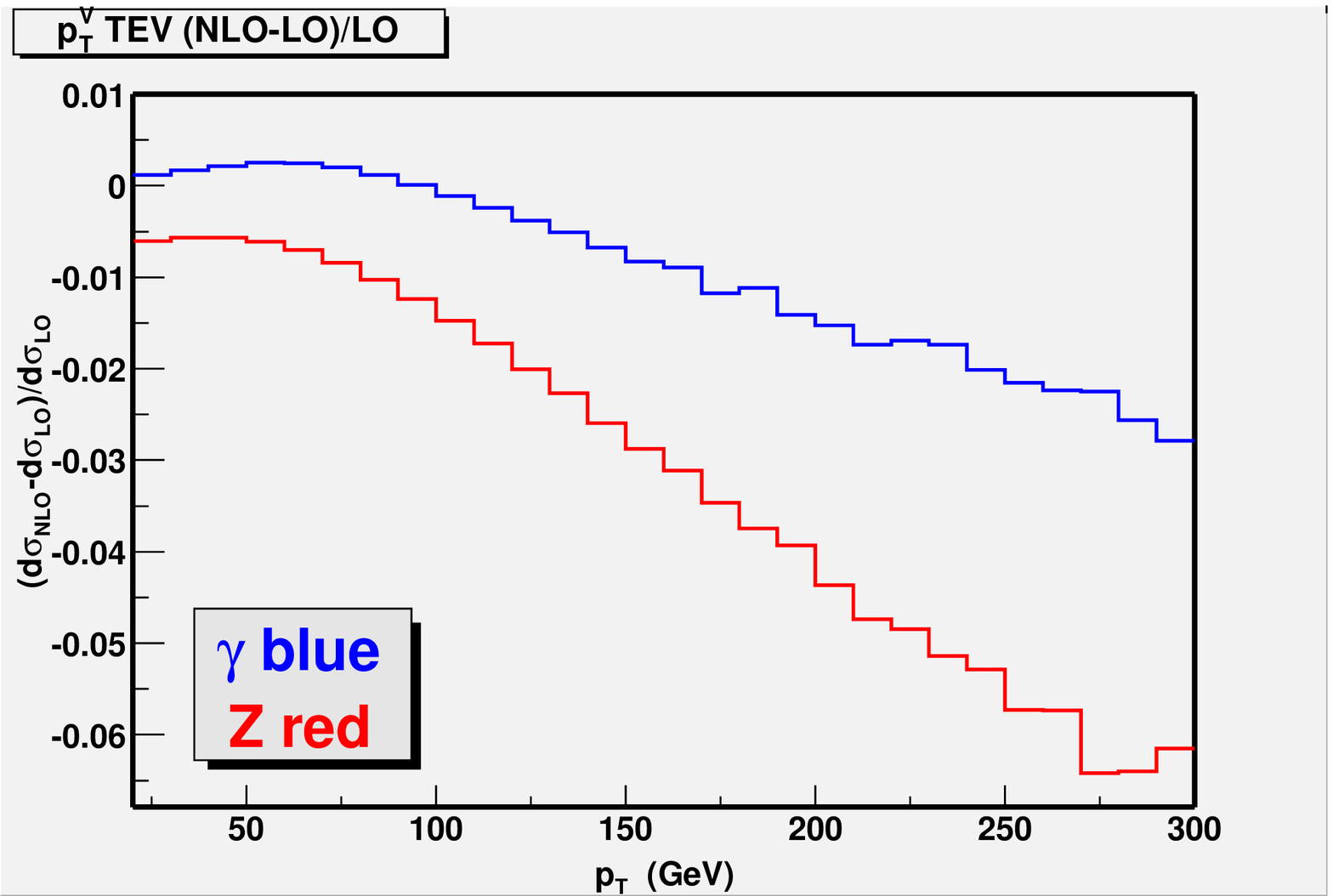, width=15cm, angle=0}}
\end{center}
\caption{\small The transverse momentum dependence of the $\gamma$- and
$Z$-boson cross 
sections in $q\bar q \to g V\quad{\rm{and}}\quad q(\bar q) g\to q(\bar q) V$
at LO 
(top frame) and the size of the one-loop weak 
corrections (bottom frame), at Tevatron ($\sqrt s_{p\bar p}=2$ TeV).
Notice that the pseudorapidity range of the jet
in the final state is limited to $|\eta|<3$.} 
\label{fig:pT-Tev}
\end{figure}